\documentclass[prd,twocolumn,eqsecnum,showpacs,nofootinbib]{revtex4}
\usepackage{bm}
\usepackage{amssymb}
\usepackage{graphicx}
\usepackage{epstopdf}

\begin{document}
\title{Self--force gravitational waveforms for extreme and intermediate mass ratio inspirals. III: Spin--orbit coupling revisited}
\author{Lior M.~Burko$^{1,2}$ and Gaurav Khanna$^3$}
\affiliation{$^1$ School of Science and Technology, Georgia Gwinnett College, Lawrenceville, Georgia 30043 \\
$^2$ Department of Physics, Chemistry, and Mathematics, Alabama A\&M University, Normal, Alabama 35762\\
$^3$ Department of Physics, University of Massachusetts, Dartmouth, Massachusetts  02747}
\date{March 17, 2015. Revised April 30, 2015}
\begin{abstract} 
The first-- and second--order dissipative self force and the first order conservative self force are applied together with spin--orbit coupling to the quasi--circular motion of a test mass in the spacetime of a Schwarzschild black hole, for extreme or intermediate mass ratios. The partial dephasing  of the gravitational waveform (at the order that is independent of the system's mass ratio) due to the self force is compared with that of spin--orbit coupling. We find that accurate waveforms for parameter estimation need to include both effects. Specifically, we find a particular value for the spin parameter such that the spin--orbit effect cancels out the self--force effect on the waveform. Exclusion of dephasing effects that are independent of the mass ratio therefore might lead to a non--perturbative error in the estimation of the system's parameters. 

\end{abstract}
\maketitle

\section{Introduction }

When a small compact object moves in the spacetime of a much more massive black hole, the orbit of the former decays and gravitational waves are emitted. The emitted waves can be observed by spaceborne observatories such as eLISA, which was chosen as the L3 mission within the European Space Agency's Cosmic Vision Program, tentatively scheduled for launch in 2034.   

The evolution of the phase $\Phi$ of the gravitational waves may be expanded in perturbation theory in the mass ratio $\varepsilon:=\mu/M$ of the system, where $\mu$ is the mass of the small companion, and $M$ is the mass of the central black hole, such that $\mu\ll M$. Specifically, we write \cite{flanagan-hinderer}, 
\begin{equation}\label{pert}
\Phi=\varepsilon^{-1}\left[\,\Phi^{(0)}+\varepsilon \, \Phi^{(1)}+O(\varepsilon^2)\right]\, .
\end{equation}
Here, we are focusing on quasi--circular orbits around a Schwarzschild black hole, which allow us to separate clearly the dissipative and conservative effects. The leading order term, $\Phi^{(0)}$, is contributed by the dissipative self force (SF) at $O(\varepsilon^2)$, when the SF is expanded in powers of $\varepsilon$. (See \cite{burko-khanna_2} for the expansion scheme.) This first--order effect is well understood \cite{poisson,barack-2009}. The next term, $\Phi^{(1)}$, includes, in addition to contributions from the dissipative SF at $O(\varepsilon^3)$ and the conservative SF at $O(\varepsilon^2)$, also contributions from spin--orbit coupling \cite{burko-2004}. The last decade has seen much progress in the theoretical understanding of self forces in General Relativity, and also significant advances in methods for their computation. For recent reviews see \cite{poisson,barack-2009,wardell}. 

A non--comprehensive list of previous work on related topics include the adiabatic evolution of extreme mass ratio inspirals  \cite{Glampedakis-2002}, 
adiabatic evolution including self--force effects \cite{Drasco-2005}, the study of the orbital evolution in eccentric binaries of non--rotating black holes \cite{Warburton-2012}, and the self-consistent evolution for a scalar charge in Schwarzschild \cite{Diener-2012}. Spin effects on gravitating systems were considered in \cite{Papapetrou-1951}, the gravitational self-torque and spin precession in \cite{Dolan-2014}, 
precession dynamics in numerical relativity and post--Newtonian approximation in \cite{Ossokine-2015}, and within the context of the 
effective one body approach in \cite{Taracchini-2014}. Self--force effects on the gravitational waveforms contributing to $\Phi^{(1)}$ within the post--Newtonian approach were considered in \cite{Isoyama-2013}, and second--order self forces were studied in 
Refs.~\cite{Pound-2012,Gralla-2012}.

The quasi--circular Schwarzschild orbits on whose evolution we are focusing are orbits that would be circular in the absence of orbital degradation. Specifically, the SF expressions that we use are those derived for circular orbits \cite{barack-sago}. We therefore introduce an error that results in an additional dissipation at $O(\varepsilon^3)$ that also contributes to $\Phi^{(1)}$. To obtain the waveforms at $O(\epsilon ^0)$ one would be required to include all contribution to $\Phi^{(1)}$. Here, however, we are interested in a more limited goal, specifically, study the relative importance of the various terms, particularly the partial dephasing due to spin--orbit coupling compared with the partial dephasing due to the dissipative SF at $O(\varepsilon^3)$ and the conservative SF at $O(\varepsilon^2)$. We therefore may neglect the partial dephasing due to our ignorance of the corrections to the SF because of the actual quasi--circular shape of the orbit. 

The study of SF effects together with spin--orbit coupling was undertaken in Ref.~\cite{burko-2004}. However, \cite{burko-2004} is limited in two important senses: First, the orbital evolution is obtained by a leading--order perturbation expansion; second, the SF needed for the computation of the orbital evolution was as yet unknown. The SF was modeled in \cite{burko-2004} to be proportional to its scalar field counterpart. Consequently, the results of \cite{burko-2004} are not quantitatively accurate. 

In Paper I \cite{lackeos_burko_12} it was shown that the dephasing $\Phi^{(1)}$ of the gravitational waveform (at $O(\varepsilon^0)$) due to the first--order conservative piece of the SF at $O(\varepsilon^2)$ is important for gravitational wave detection and astronomy. In Paper II \cite{burko-khanna_2} we considered the relative importance to the dephasing of the second--order dissipative piece of the SF (at $O(\varepsilon^3)$). Specifically, it was shown in Paper II that for the class of orbits studied there the partial dephasing due to the second--order dissipative SF (at $O(\varepsilon^3)$) amounts to $8\%$ of the partial dephasing due to the first--order conservative SF (at $O(\varepsilon^2)$), even though they both contribute to gravitational wave dephasing at $O(\varepsilon^0)$. 
In this paper we add the effect of spin--orbit coupling, and study, within the testbed of quasi-circular Schwarzschild orbits, the relative partial dephasing due to all of the above effects that contribute to dephasing of the gravitational waveform at $O(\varepsilon^0)$. The introductions to papers I and II supplement this introduction, and the reader is referred to further detail therein. 
The organization of this paper is as follows: In Section II we describe the physical model of the system that we study, and discuss the assumptions made in the modeling of the SF and the spin--orbit coupling. Section III discusses the findings of our numerical experiments. First, we discuss the properties of the gravitational waveforms, and then the partial dephasing of the gravitational waveforms due to the effects under consideration. 

\section{The Physical Model}

We wish to study the relative importance of the partial dephasing due to spin--orbit coupling compared with that of the SF, specifically the first--order conservative SF (at $O(\varepsilon^2)$) and the second--order dissipative SF (at $O(\varepsilon^3)$). We therefore specialize to the simplest system that allows us to compare the various partial dephasing effects. Specifically, we consider quasi-circular Schwarzschild orbits of a spinning point mass, where the spin is aligned (or anti--aligned) with the orbital angular momentum. Denoting the Pauli--Lubanski spin pseudovector by $S^{\alpha}$, the Papapetrou equations \cite{Papapetrou-1951} imply that $S^{\alpha}=s\mu^2/r\,\delta^{\alpha}_{\theta}$ where $-1\leq s\leq 1$ for a mass representing a black hole (see \cite{burko-2004} for more detail), and the equations of motion become 
\begin{equation}
\mu\,\frac{\,Du^t}{\,d\tau}=f_{\rm SF}^t+3s\frac{M\mu^2}{r^2}\left(1-\frac{2M}{r}\right)^{-1}u^{\varphi}u^r
\end{equation}
\begin{equation}
\mu\,\frac{\,Du^r}{\,d\tau}=f_{\rm SF}^r+3s\frac{M\mu^2}{r^2}\left(1-\frac{2M}{r}\right)u^{\varphi}u^t
\end{equation}
\begin{equation}
\mu\,\frac{\,Du^\varphi}{\,d\tau}=0\, ,
\end{equation}
where $u^{\alpha}$ is the four--velocity of the mass $\mu$, $\tau$ is its proper time, $r$ is the radial Schwarzschild coordinate, and the operator $D/\,d\tau$ denotes covariant differentiation with respect to proper time compatible with the background Schwarzschild metric. Notice that $s$ is a free parameter, which we will vary in our numerical experiment. When $s=0$ the problem is reduced to the one studied in Paper II. 
The temporal component of the SF, $f_{\rm SF}^t$, includes both contributions at $O(\varepsilon^2)$ and at $O(\varepsilon^3)$. The radial component,  $f_{\rm SF}^r$, includes contributions at $O(\varepsilon^2)$. For the SF we use the expressions listed in the Appendix of Paper II.  Specifically, the dissipative SF at $O(\varepsilon^2)$ and the conservative SF at the same order are fully relativistic and were derived in the Lorenz gauge  \cite{barack-sago}. The dissipative SF at $O(\varepsilon^3)$ is approximated by its 3.5 post--Newtonian expansion, in the harmonic gauge. The SF is therefore obtained in a hybrid gauge. At $O(\varepsilon^2)$ the Lorenz gauge and the harmonic gauge give results that agree to the numerical accuracy of our computation. We therefore propose that our SF is therefore written in a consistent harmonic gauge to the numerical accuracy of our computation.  

\begin{figure}
 \includegraphics[width=7.5cm]{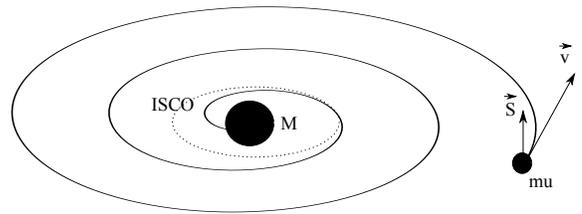}
\caption{Schematic representation (not to scale) of our physical model: a small compact mass $\mu$ with spin angular momentum $S^{\alpha}=s\mu^2/r\,\delta^{\alpha}_{\theta}$ which is aligned (as in this drawing) or anti--aligned with the orbital  angular momentum, orbits with three-velocity $\vec{v}$  a Schwarzschild black hole of mass $M$ in a quasi--circular orbit. The latter decays because of the combined effects of radiation reaction and spin--orbit coupling, a trajectory represented by the solid spiral. The ISCO is represented by a dotted circle.}
\label{config}
\end{figure}

The set up of our physical model is represented schematically in Fig.~\ref{config}. 

This model is particularly simple not just because the Pauli--Lubanski pseudovector is known exactly, but primarily because the introduction of the spin does not alter the planar nature of the mass's trajectory. We may therefore apply the same computational method that we applied in Paper II, and the reader is referred there for detail. Our computational approach is based on using the method of osculating geodesics \cite{Pound-Poisson-2008} for the orbital evolution, and on solving the Teukolsky equation for a source based on the resulting orbit for finding the gravitational waveforms. 
As the terms that depend on the particle's spin in the source term for the Teukolsky equation are smaller by a factor of $\varepsilon$ than the corresponding terms that are independent of the particle's spin, the former may be neglected. In practice, the numerical experiments reported on here use $\varepsilon=10^{-2}$. Therefore, our results for the dephasing are accurate to $1\%$. Neglecting the spin--coupled terms in the source term for the Teukolsky equation shortens the computation considerably. Indeed, tests that we have performed show that the actual error that is introduced in the waveform's phase is indeed of the magnitude predicted. 

The effect of interest is independent of $\varepsilon$, a conclusion which was tested in detail in Papers I and II. We may therefore shorten the computation time significantly by studying a rather large value for $\varepsilon$, in practice $10^{-2}$. We take in practice the orbit to start at $r=10M$, and follow the evolution down to the Innermost Stable Circular Orbit (ISCO) at $r=6M$. The orbital evolution of $\sim 50$ orbits last about $\sim9\times 10^3M$. The duration of the motion, and the number of orbits done vary depending on the value of $s$ chosen, as detailed below. We show in practice results for the $h_{22}$ mode of the field. 

\section{Numerical results}

\subsection{The waveforms}
In all cases, we start with waveforms that at $t=0$ are in phase. We then consider the wave train, and monitor the relative dephasing. In practice, we pose the following thought experiment: Consider a number of isolated astrophysical systems, each consisting of a binary made of a Schwarzschild black hole and a spinning black hole in quasi--circular orbit, with mass ratio $\varepsilon=10^{-2}$. All the spin angular momenta are aligned or anti--aligned with the orbital angular momentum. The various systems are different in the spin angular momentum of the smaller black hole, which we vary in the range $-1\le s\le 1$. In practice, we vary $s$ in increments of $\,\Delta s=0.2$.

\begin{figure}
 \includegraphics[width=7.5cm]{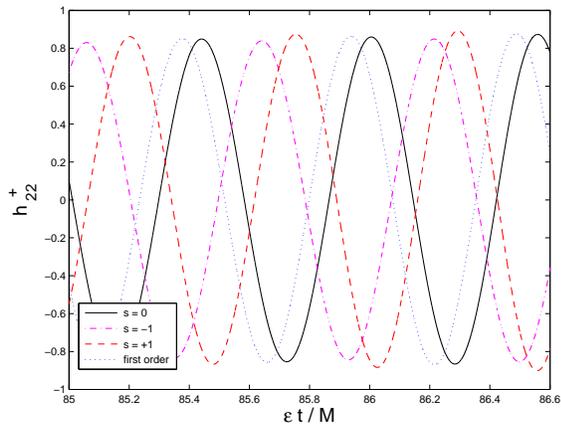}
\caption{The polarization state $h^+_{22}$ of the gravitational waveforms as a function of the time. Shown are the cases $s=0$ (solid curve), $s=-1$ (dash--dotted curve), $s=+1$ (dashed curve), and $\Phi^{(1)}\equiv 0$ (dotted curve). All these cases are in phase at $t=0$. }
\label{wf}
\end{figure}

\begin{figure}
 \includegraphics[width=7.5cm]{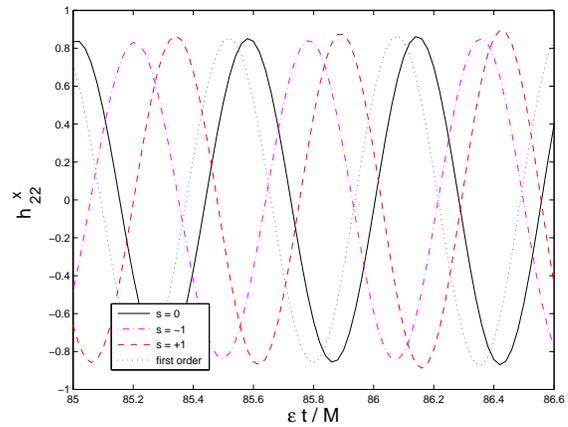}
\caption{Same as Fig.~\ref{wf} for the $h^{\times}_{22}$ polarizations state.}
\label{wfc}
\end{figure}

In Figs. \ref{wf} and \ref{wfc} we show the $h^+_{22}$ and $h^{\times}_{22}$ polarizations states, respectively, for the cases $s=-1,0,1$ and also for the case that the second order effects in the waveforms are turned off completely (i.e., with no second--order dissipative effect, no first--order conservative effect, and no spin--orbit coupling effect, or equivalently $\Phi^{(1)}\equiv 0$). We find that substantial phase difference builds up between these cases. Specifically, at a given value of time, the phase evolution of $s=+1$ is greater than than of $s=0$, and that of $s=-1$ is smaller than that of $s=0$. The first--order waveform ($\Phi^{(1)}\equiv 0$) has phase evolution in between that of $s=0$ and $s=+1$. We quantify these statements and make them precise below, in Sec.~\ref{dephasing}

\begin{figure}
 \includegraphics[width=7.5cm]{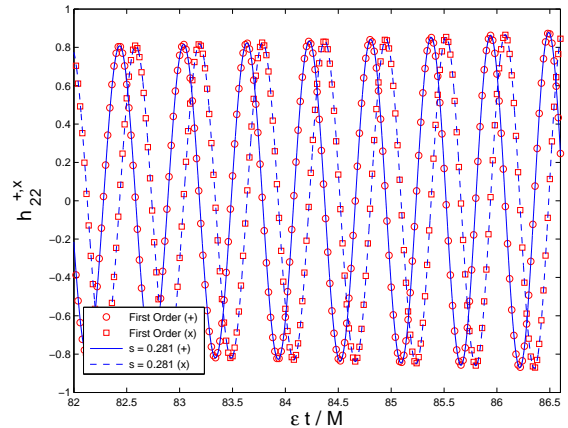}
\caption{The waveforms for the two polarization states $h^{+,\times}_{22}$ for the case $s=0.281$ (solid ($+$) and dashed ($\times$) curves), and for the $\Phi^{(1)}\equiv 0$ case ($\circ$ ($+$) and $\square$ ($\times$) as functions of the time $t$.) For each polarization state, the two waveforms are in phase at $t=0$. }
\label{wfs0281}
\end{figure}

We next raise the following question: can one find a value for $s$ such that the second--order waveform will overlap the first--order ($\Phi^{(1)}\equiv 0$)) waveform? The scenario we propose is as follows: assume the actual data stream at the detector is modeled by the second--order waveform. We then use the first--order waveform as a template. Is there a source for which the first--order waveform would give us a good {\em global} match, and lead us to an estimation of the system's parameters that is not just perturbatively inaccurate, but incorrect by a large value of the parameters?

\begin{figure}
 \includegraphics[width=7.5cm]{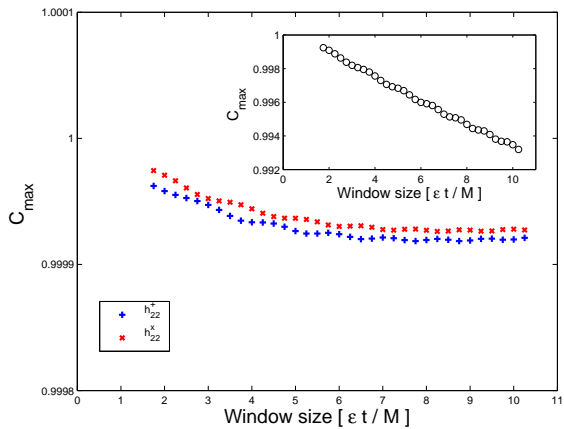}
\caption{The overlap integral $C_{\rm max}$ for either polarization states $h^{+,\times}_{22}$ for the case $s=0.281$ (represented by $+$ and 
$\times$ correspondingly) with the corresponding $\Phi^{(1)}\equiv 0$ case as functions of the length of the interval of the segment being integrated. The inset shows the same for $h^{\times}_{22}$ (represented by $\circ$) for the case $\Phi^{(1)}\equiv 0$ and $h^{+}_{22}$ for the case $s=0.281$.}
\label{overlap_integral_fo_p0281}
\end{figure}

In Fig.~\ref{wfs0281} we show the waveforms for $h^{+,\times}_{22}$ for the case $s=0.281$ (see below in Sec.~\ref{dephasing} for motivation for studying this particular value for $s$) and for the case $\Phi^{(1)}\equiv 0$ (the latter case implies zero spin--orbit coupling in particular). The two sets of waveforms appear to be almost exactly in phase with each other, as is shown in Fig.~\ref{overlap_integral_fo_p0281}, which plots the overlap integral $C_{\rm max}$ for either polarization state of $h_{22}$ with the first--order waveform ($\Phi^{(1)}\equiv 0$). The overlap integral is calculated as 
\begin{equation}
C_{\rm max}={\rm max}_{\tau}\frac{< h_{22}^1(t) | h_{22}^2(t-\tau) >}{\sqrt{< h_{22}^1(t) | h_{22}^1(t) >\; < h_{22}^2(t) | h_{22}^2(t) >}}
\end{equation}
(see more detail in Paper I). Either overlap integral changes very little as a function of the integration interval (``window size"), and for all intervals tested was smaller than unity by less than $10^{-4}$ (the longest interval we show corresponds to $\sim 36$ orbits). Since the two waveforms are almost indistinguishable, we may raise the question of whether we can identify the polarization state correctly, even when our parameter estimation is off by a large amount. The inset in Fig.~\ref{overlap_integral_fo_p0281} shows the overlap integral as a function of the integration interval for the template waveform in one polarization state, and the data stream in the other polarization state. The data show that indeed one could distinguish the two polarization states even in such a case.

\subsection{Waveform Dephasing}\label{dephasing}

\begin{figure}
 \includegraphics[width=7.5cm]{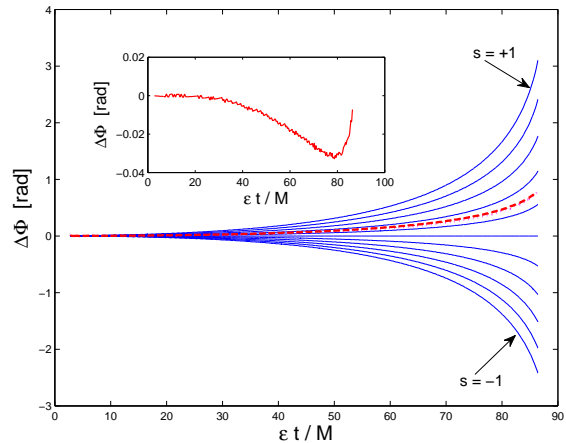}
\caption{The evolution of the relative phase of the gravitational waveform between the second--order waveform with $s\ne 0$ and the case $s=0$ (solid (blue) curves) in increments of $\,\Delta s=0.2$, and between the first--order waveform ($\Phi^{(1)}\equiv 0$) and the second--order waveform with $s=0$ (dashed (red) curve). The inset shows the same between the first--order waveform ($\Phi^{(1)}\equiv 0$) and the second--order waveform with $s= 0.281$. On this scale the two polarization states look indistinguishable.}
\label{dephase_evol}
\end{figure}

We next study the partial dephasing of the waveforms due to the spin--orbit coupling effect. In Fig.~\ref{dephase_evol} we show the evolution of the dephasing of the gravitational wave for various values of $s$. The range of values for the dephasing is $\,\Delta\phi\in [-2.409,3.089]\,{\rm rad}$ for the range $s\in [-1,1]$. In comparison, the magnitude of the second--order effect without spin--orbit coupling is $0.797\,{\rm rad}$, the latter being contributed mostly from the first--order conservative effect (and to a smaller extent by the second--order dissipative effect). 

The inset in  Fig.~\ref{dephase_evol} shows the dephasing between the first--order waveform ($\Phi^{(1)}\equiv 0$) and the second--order waveform with $s=0.281$. This dephasing is nowhere along the evolution greater than $|\Delta\Phi|\sim 0.03\,{\rm rad}$, which makes them practically nearly indistinguishable from the observational point of view. This conclusion complements the observations made based on the overlap integrals as shown in Fig.~\ref{overlap_integral_fo_p0281}. In practice, the implication is that if one were to neglect the dephasing at $O(\varepsilon^0)$, i.e., the term $\Phi^{(1)}$, one would make a non--perturbative error in the estimation of the system's parameters. Specifically, if one models the actual data stream for the waveform for the case $s=0.281$ with the waveform including the $\Phi^{(1)}$ term, and uses a template that neglects the $\Phi^{(1)}$ term for the case $s=0$, the cross correlation of the data stream with the template would indicate nearly exact matching, even though the spin parameter of the actual system is non--perturbatively different than the spin parameter used for the template. 

\begin{figure}
 \includegraphics[width=7.5cm]{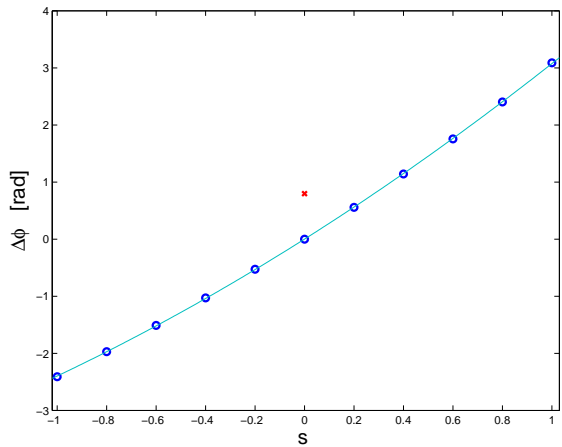}
\caption{The relative phase of the gravitational waveform between the second--order waveform with $s\ne 0$ and the case $s=0$ ($\circ$ (blue) and the corresponding best fit curve), and between the first--order waveform ($\Phi^{(1)}\equiv 0$) and the second--order waveform with $s=0$ ($*$ (red)). The values of $s$ shown are between $-1$ and $+1$ in increments of $0.2$). All data are taken at $\varepsilon t/M=86.40$. The fit function is the parabola $\,\Delta\phi=0.33817\,s^2+2.7367\, s+9.1375\times 10^{-4}$. On this scale the two polarization states look indistinguishable.}
\label{phases}
\end{figure}

We next consider the total dephasing $\,\Delta\Phi$ as a function of the spin parameter $s$ over a fixed interval of time, comparing the dephasing between the second--order waveform with $s\ne 0$ and with $s=0$. Figure \ref{phases} suggests a quadratic dependence of $\,\Delta\Phi$  on $s$. Figure \ref{phases} also shows the dephasing during the same time interval between the first--order waveform  ($\Phi^{(1)}\equiv 0$) and the second--order waveform with $s=0$ (0.797 rad). Equating the best fit quadratic function for the former to the latter value, we can solve for the $s$ value that would produce the same dephasing. The dephasing in that case is found when the spin parameter $s=0.281$. We find this value when we compare the final, cumulative dephasing at the end of the entire wave train. (The other solution corresponds to an $s$ value outside the allowed range for black holes. However, if the compact object is a Neutron star, the other $s$ value may be relevant, and present two possible $s$ values, that may both enrich and complicate the parameter estimation problem.)  
We have shown above, nevertheless, that the two corresponding waveforms overlap almost precisely over the entire orbital evolution, and the two phases nowhere are different by more than $\sim 0.03$ rad.  We leave for future study the question of how this special value of $s$ depends on the system's parameters, specifically on the initial size of the orbit.

\begin{figure}
 \includegraphics[width=7.5cm]{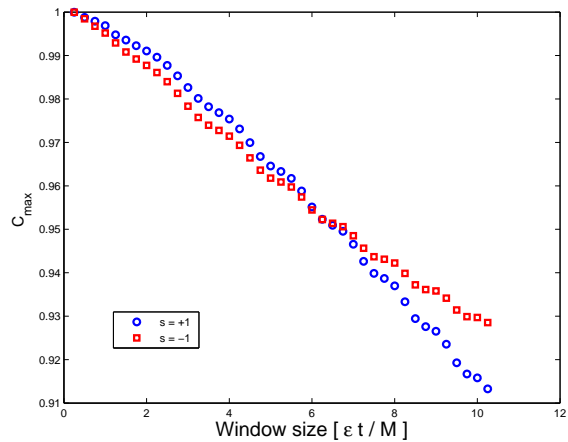}
\caption{The value of $C_{\rm max}$ as a function of the duration of the integration interval for the polarization state $h_{22}^+$. Shown are the cases $s=+1$ ($\circ$, blue) and $s=-1$ ($\square$, red). In both cases the overlap integral is calculated with the waveform for $s=0$.}
\label{cmax_m1p1}
\end{figure}

The dephasing of the waveform because of the spin--orbit coupling effect is shown in Fig.~\ref{cmax_m1p1}, which shows the overlap integral $C_{\rm max}$ as a function of the duration of the integration interval (the ``window size'') for two values of the spin parameter, specifically $s=+1$ and $s=-1$. As expected, the longer the integration interval, the smaller the overlap integral. 

\begin{figure}
 \includegraphics[width=7.5cm]{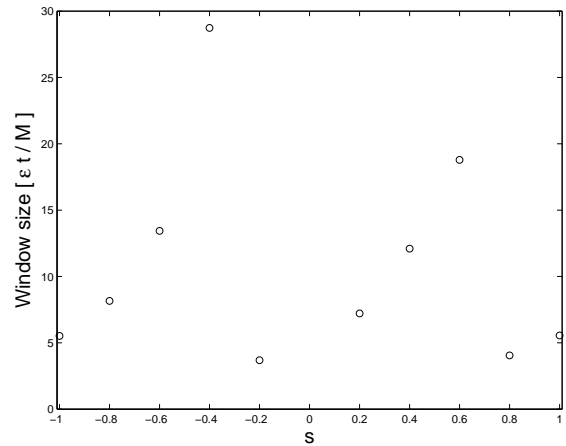}
\caption{The length of the window for which the overlap integral $C_{\rm max}=0.96$ as a function of the spin parameter $s$ for the polarization state $h_{22}^+$. }
\label{c096}
\end{figure}

One may ask how long does the integration interval need to be for the overlap ingress to drop below $C_{\rm max}^{\rm crit}=0.96$ for which the detection rate drops by $10\%$. Figure \ref{c096} shows the length of the required integration interval as a function of the spin parameter $s$. The data points suggest a bimodal curve. The data point for $s=0$ is not shown, as that case corresponds for the overlap integral of the $s=0$ wafers with itself.

\section*{Acknowledgements} 
L.M.B.~acknowledges research support from NSF Grant No.~DUE--1300717.  
G.K.~acknowledges research support from NSF Grants No.~PHY--1303724 and No.~PHY--1414440, and from the U.S.~Air Force agreement No.~10--RI--CRADA--09.

\end{document}